\documentclass[12pt,a4paper]{article}

\usepackage{amsmath, amssymb}
\usepackage{graphicx}
\usepackage{pstricks}
\usepackage{pst-plot}

\begin{document}

\title{Time symmetric electrodynamics, electric charge conservation, and the Lorenz gauge}
\author{
C\u alin Galeriu \\
%Mark Twain International School, Voluntari, IF, Romania \\
%calin.galeriu@marktwainschool.ro
}

\maketitle

\section*{Abstract}

{\it We reveal a new way in which the Lorenz gauge condition 
is related to the electric charge conservation,
in a universe where electrically charged point particles are created and annihilated.
We derive our results using time symmetric electrodynamics, 
relying on observations made
by Jacov Frenkel (in 1925) and John Archibald Wheeler (in 1940).
Both the Lorenz gauge condition and the electric charge conservation
are expressed as closed path integrals in Minkowski space. The integration path
is the same, it is the sum of all \lq\lq Wheeler electrons\rq\rq .}

{\bf Keywords} one-electron universe, action-at-a-distance, continuity equation

\section{Introduction}

There are many ways in which Maxwell's equations can be solved in order to obtain the
electromagnetic four-potential of a point particle. Most often \cite{Jackson1999}
one starts by imposing the Lorenz gauge condition
\begin{equation}
\partial_\alpha A^\alpha = 0,
\label{eq:Lorenz_gauge}
\end{equation}
where the four-potential $A^\alpha$ is $(\Phi, {\bf A})$, 
in Gaussian units,
and the covariant differential operator $\partial_\alpha$ is 
$(\frac{1}{c}\frac{\partial}{\partial t}, {\bf \nabla})$.
In this way we obtain four decoupled wave equations, one for each component of the four-potential
\begin{equation}
\Box A^\alpha = \frac{4 \pi}{c} J^\alpha,
\label{eq:wave_equation}
\end{equation}
where the four-current density $J^\alpha$ is $(c \rho, {\bf J})$,
and the d'Alembertian operator $\Box$ is 
$\partial_\alpha \partial^\alpha = \eta^{\alpha \beta} \partial_\alpha \partial_\beta$.
The Minkowski metric tensor $\eta^{\alpha \beta}$ has signature $(1, -1, -1, -1)$.
The four-current density satisfies the continuity equation
\begin{equation}
\partial_\alpha J^\alpha = 0.
\label{eq:continuity_eq}
\end{equation}

The conservation of electric charge is an empirical fact that does not need a mathematical 
demonstration. In our calculations, the conservation of electric charge is taken for granted.
Nonetheless we notice that, once Maxwell's equations are accepted,
one can derive the continuity equation. 
This only shows that Maxwell's equations are consistent with the continuity equation.

The Lorenz gauge condition is an assumption that we make in order to derive the 
Li\'{e}nard-Wiechert electromagnetic potentials. Since this gauge is our choice,
and we are free to chose differently, we cannot talk about a mathematical proof of it.
Nonetheless we notice that, once the Li\'{e}nard-Wiechert potentials are accepted,
one can derive the Lorenz gauge condition. 
This only shows that the Li\'{e}nard-Wiechert potentials are consistent with the Lorenz gauge.

Any conservation law can be written
in a differential form, as in (\ref{eq:continuity_eq}), 
or in an integral form,
in which
the rate of variation of the 3D volume integral of the relevant density
is related to the 2D closed surface integral of the corresponding flux.

Frenkel \cite{Frenkel1925} has also expressed the Lorenz gauge condition as a 1D integral over a 
closed path in the complex plane. We give in Section 2 a modern account of his theory.

Inspired by Frenkel's idea, 
and using time symmetric electrodynamics,
in Section 3 we express the Lorenz gauge condition 
as a 1D integral over a closed path in Minkowski space. In fact this integration path is a sum of
very special closed paths, each of these individual closed paths being,
by definition, a \lq\lq Wheeler electron\rq\rq.
We also express the
continuity equation as a 1D integral over the same closed path. 
This common integration path 
is the newly discovered connection between the Lorenz gauge condition and the
electric charge conservation.

\section{A modern exposition of Frenkel's work}

The inhomogeneous electromagnetic wave equation (\ref{eq:wave_equation}) 
can be solved by finding its Green function, which satisfies the equation
\begin{equation}
\Box_x G(x, x') = \delta^{4D}(x - x'),
\label{eq:Green_fn_equation}
\end{equation}
where $x = (c t, {\bf x})$ and $x' = (c t', {\bf x'})$ are position four-vectors in Minkowski space. 
In the absence of boundary surfaces, the four-potential is given by 
\begin{equation}
A^\alpha (x) = \frac{4 \pi}{c} \int G(x, x') \, J^\alpha (x') \, d^4x'.
\label{eq:Green_fn_solution}
\end{equation}

At this moment, instead of solving equation (\ref{eq:Green_fn_equation}) 
in order to find $G(x, x')$, we simply notice that \cite{Anderson1967, Ryder1974}
\begin{equation}
\Box \delta(x^2) = 4 \pi \delta^{4D}(x),
\label{eq:Dirac_delta_property_1}
\end{equation}
which allows us to identify the Green function as
\begin{equation}
G(x, x') = \frac{1}{4 \pi} \delta\big( (x - x')^2 \big). 
\label{eq:symmetric_Green_fn}
\end{equation}

The electromagnetic four-potential (\ref{eq:Green_fn_solution}) becomes
\begin{equation}
A^\alpha (x) = \frac{1}{c} \int \delta\big( (x - x')^2 \big) \, J^\alpha (x') \, d^4x'.
\label{eq:four_potential}
\end{equation}

To this solution of equation (\ref{eq:wave_equation}) one could also add a solution of the 
homogeneous electromagnetic wave equation. When working with only the 
retarded (advanced) potential, this incoming (outgoing) field
takes care of the initial (final) conditions \cite{Jackson1999}.
When working with time symmetric action-at-a-distance electrodynamics,
there is no need for such incoming or outgoing fields \cite{Anderson1967,Wheeler1949}.

While it is common practice to use another Dirac delta function identity
\begin{equation}
\delta(w^2 - a^2) = \frac{1}{2 |a|} \big[ \delta(w - a) + \delta(w + a) \big],
\label{eq:Dirac_delta_property_2}
\end{equation}
in order to split the Green function (\ref{eq:symmetric_Green_fn}) into causal and acausal parts,
and the time symmetric potential (\ref{eq:four_potential}) into retarded and advanced parts,
such course of action would not be helpful to our next derivation steps.

We now turn our attention toward the four-current density. 
Consider an electrically charged point particle $A$,
with an electric charge $q_A$,
whose worldline is given by 
$x_A(t) = \big( c t, {\bf x_A}(t) \big)$. 
The velocity of this particle is ${\bf v_A}(t) = \frac{d {\bf x_A}}{d t}$.
The electric charge density is given by
\begin{equation}
\rho({\bf x'}, t') = q_A \, \delta^{3D}\big( {\bf x'} - {\bf x_A}(t') \big),
\label{eq:charge_density}
\end{equation}
while the electric current density is given by
\begin{equation}
{\bf J}({\bf x'}, t') = q_A \, {\bf v_A}(t') \, \delta^{3D}\big( {\bf x'} - {\bf x_A}(t') \big).
\label{eq:current_density}
\end{equation}

The four-current density can be written in a manifestly covariant form by introducing an additional
Dirac delta function $\delta(t' - t) = c \, \delta(c t' - c t)$
and by integrating over the time, obtaining \cite{Jackson1999}
\begin{multline}
J^\alpha (x') = q_A \, U_A^\alpha(t')  \, 
\delta^{3D}\big( {\bf x'} - {\bf x_A}(t') \big) \frac{1}{\gamma_A(t')} \\
= \int q_A \, U_A^\alpha(t)  \, 
\delta^{3D}\big( {\bf x'} - {\bf x_A}(t) \big) \frac{1}{\gamma_A(t)} \delta(t' - t) dt \\
= \int q_A \, U_A^\alpha(t) \, \delta^{4D}\big( x' - x_A(t) \big) \frac{c}{\gamma_A(t)} dt \\
= \int q_A \, U_A^\alpha(\tau_A) \, \delta^{4D}\big( x' - x_A(\tau_A) \big) c \, d\tau_A,
\label{eq:four_current_density}
\end{multline}
where $U_A^\alpha = (\gamma_A \, c, \gamma_A \, {\bf v_A})$ is the four-velocity, 
$\gamma_A$ is the corresponding Lorentz factor, and 
$\tau_A$ is the proper time of particle $A$.

It is generally assumed that it doesn't matter what happens with the worldline of
particle $A$ very early, before it intersects the past lightcone (with vertex at the field point),
or what happens with the worldline very late, after it intersects the future lightcone.
This is because only the retarded electromagnetic potentials, and maybe also the 
advanced potentials, are needed in order to calculate the electromagnetic field at the point of interest. 
As a consequence, sometimes the limits of integration in equation (\ref{eq:four_current_density}) 
are left unspecified, 
like for example in \cite{Jackson1999}, and sometimes the integral is extending from $- \infty$
to $+ \infty$, like for example in \cite{Anderson1967,Wheeler1949,Thirring1958,Chubykalo2016}. 
The later option is in direct conflict with
the assumption that \lq\lq the sources are localized in space and time\rq\rq \cite{Jackson1999}.
To avoid any ambiguity, we should recognize that particle $A$ is created
at the initial time $t_A^{(i)}$ and annihilated at the final time $t_A^{(f)}$, 
and we should explicitly show these actual limits of integration in equation (\ref{eq:four_current_density}).
We will return to this idea later.

With the four-current density (\ref{eq:four_current_density}), 
the four-potential (\ref{eq:four_potential}) becomes
\begin{equation}
A^\alpha (x) = \frac{1}{c} \int \delta\big( (x - x')^2 \big) \, 
\Big( \int q_A \, U_A^\alpha(\tau_A) \, \delta^{4D}\big( x' - x_A(\tau_A) \big) c \, d\tau_A \Big)
\, d^4x'.
\label{eq:Green_fn_solution_2}
\end{equation}

We perform the four-dimensional integration and we obtain
\begin{equation}
A^\alpha (x) = \int \delta\Big( \big(x - x_A(\tau_A) \big)^2 \Big) \, 
q_A \, U_A^\alpha(\tau_A) \, d\tau_A.
\label{eq:Green_fn_solution_3}
\end{equation}

Frenkel \cite{Frenkel1925}, working before the time when Dirac \cite{Dirac1927} introduced his delta function, 
and using more heuristic arguments, reaches a similar conclusion. Following Frenkel,
we calculate the divergence of the four-potential (\ref{eq:Green_fn_solution_3}) and we obtain
a perfect differential inside the integral
\begin{multline}
\frac{\partial A^\alpha (x)}{\partial x^\alpha} 
= \int \Big[ \frac{\partial}{\partial x^\alpha} 
\delta\Big( \big(x - x_A(\tau_A) \big)^2 \Big) \Big] \, q_A \, U_A^\alpha(\tau_A) \, d\tau_A \\
= - \int \Big[ \frac{\partial}{\partial x_A^\alpha} 
\delta\Big( \big(x - x_A(\tau_A) \big)^2 \Big) \Big] \, q_A \, \frac{dx_A^\alpha}{d\tau_A} \, d\tau_A \\
= - \int \Big[ \frac{d}{d\tau_A} 
\delta\Big( \big(x - x_A(\tau_A) \big)^2 \Big) \Big] \, q_A \, d\tau_A.
\label{eq:divergence}
\end{multline}

At this moment Frenkel notices the fact that the Lorenz gauge condition is automatically satisfied
whenever the integral (\ref{eq:divergence}) is calculated along a closed path.
Writing before the time when the processes of creation and annihilation of particles were fully understood,
Frenkel proposes a rather abstract procedure: replace the time $t$, a real number, with a complex number,
assume that the position functions in $x_A(t)$ admit a direct analytic continuation over the entire complex plane,
and close the integration path inside this complex plane. As an added benefit, 
Cauchy's residue theorem now applies, and the Li\'{e}nard-Wiechert potentials are thus calculated. 
This derivation of the Li\'{e}nard-Wiechert potentials is also given by Stratton \cite{Stratton1941}.

\section{The Wheeler electron}

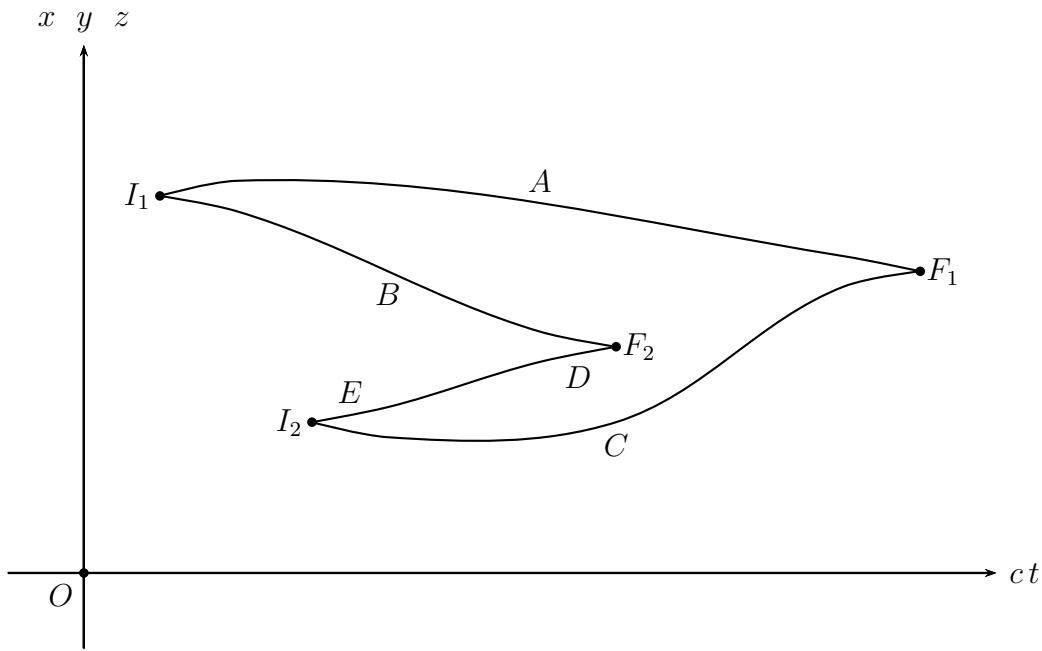
\begin{figure}[h!]
\begin{center}
\begin{pspicture}(-1,-1)(12.5,7.5)
\psaxes[ticks=none,labels=none]{->}(0,0)(-1,-1)(12,7)[$c \, t$,0][$x\ \ y\ \ z$,90]
\psdot(0,0)
\rput(-0.3,-0.3){$O$}
\pscurve(3,2)(4,2.2)(6,2.8)(7,3)
\pscurve(1,5)(2,4.8)(6,3.2)(7,3)
\pscurve(1,5)(2,5.2)(10,4.2)(11,4)
\pscurve(3,2)(4,1.8)(7,2)(10,3.8)(11,4)
\psdot(1,5)
\psdot(3,2)
\psdot(7,3)
\psdot(11,4)
\rput(6,5.2){$A$}
\rput(4,3.7){$B$}
\rput(7,1.7){$C$}
\rput(6.5,2.6){$D$}
\rput(3.5,2.4){$E$}
\rput(0.7,5){$I_1$}
\rput(11.3,4){$F_1$}
\rput(2.7,2){$I_2$}
\rput(7.3,3){$F_2$}
\end{pspicture}
\caption{A \lq\lq Wheeler electron\rq\rq \ composed of four particles.}
\label{fig:Wheeler_electron}
\end{center}
\end{figure}

We believe that the integration path in equation (\ref{eq:divergence}) can be closed in an alternative,
more intuitive procedure that keeps the time coordinate as a real number. 
Assuming that there are
no electric charges at $t \to - \infty$ and at $t \to \infty$, we must recognize that particle $A$ is
created at a time $t_A^{(i)}$ in a process that, due to the conservation of electric charge,
also produces another particle $B$ with electric charge $q_B = - q_A$. For the same reason, 
particle $A$ is annihilated at a time $t_A^{(f)}$ in a process that also eliminates another particle $C$
with electric charge $q_C = - q_A$. If particles $B$ and $C$ are the same particle, then we stop. We have 
closed the path in spacetime. If particles $B$ and $C$ are not the same particle, then we must
recognize that particle $B$ is annihilated at a time $t_B^{(f)}$ in a process that also eliminates
another particle $D$ with electric charge $q_D = - q_B = q_A$. For the same reason,
particle $C$ is created at a time $t_C^{(i)}$ in a process that also produces another particle $E$
with electric charge $q_E = - q_C = q_A$. If particles $D$ and $E$ are the same particle, then we stop. We have 
closed the path in spacetime, as shown in Figure \ref{fig:Wheeler_electron}. 
If particles $D$ and $E$ are not the same particle,
then we repeat the procedure, over and over, until the two particles coincide. This must happen sooner
or later, due to the finite number of electrically charged particles in the universe. In fact,
any electrically charged particle in the world will have to belong to 
one or another of these spacetime structures, 
its entire worldline being only a section of such a closed path. 

These closed paths, these roundtrips in spacetime, we will call them \lq\lq Wheeler electrons\rq\rq, 
for it was John Archibald Wheeler \cite{Feynman1965} who first described 
such a knot of worldlines going up and down in time, in his one-electron universe.
Now, of course, we must recognize that in reality there isn't just one very large Wheeler electron,
linking together all the electrons and all the positrons in the universe, as he originally suggested, but many of them. 
And we must also recognize that, in a given Wheeler electron, the particles with positive electric charge don't
have to necessarily be positrons, and the particles with negative electric charge don't have to
necessarily be electrons.  The process in which electrically charged particles are created could be
the birth of an electron-positron pair, but it could also be something else, like for example beta minus decay.
The process in which electrically charged particles are annihilated could be the disappearance
of an electron-positron pair, but it could also be something else, like for example electron capture.

We now go back to the formula for the four-current density (\ref{eq:four_current_density}),
and we include in it all the electrically charged particles in the universe, while at the same time 
explicitly showing the limits of integration
\begin{multline}
J^\alpha (x') 
= \int_{\tau_A^{(i)}}^{\tau_A^{(f)}} q_A \, U_A^\alpha(\tau_A) \, \delta^{4D}\big( x' - x_A(\tau_A) \big) c \, d\tau_A \\
+ \int_{\tau_B^{(i)}}^{\tau_B^{(f)}} q_B \, U_B^\alpha(\tau_B) \, \delta^{4D}\big( x' - x_B(\tau_B) \big) c \, d\tau_B \\
+ \int_{\tau_C^{(i)}}^{\tau_C^{(f)}} q_C \, U_C^\alpha(\tau_C) \, \delta^{4D}\big( x' - x_C(\tau_C) \big) c \, d\tau_C + . . . 
\label{eq:four_current_density_2}
\end{multline}

With this new formula for the four-current density, 
the four-potential becomes
\begin{multline}
A^\alpha (x) 
= \int_{\tau_A^{(i)}}^{\tau_A^{(f)}} \delta\Big( \big(x - x_A(\tau_A) \big)^2 \Big) \, 
q_A \, U_A^\alpha(\tau_A) \, d\tau_A \\
+ \int_{\tau_B^{(i)}}^{\tau_B^{(f)}} \delta\Big( \big(x - x_B(\tau_B) \big)^2 \Big) \, 
q_B \, U_B^\alpha(\tau_B) \, d\tau_B \\
+ \int_{\tau_C^{(i)}}^{\tau_C^{(f)}} \delta\Big( \big(x - x_C(\tau_C) \big)^2 \Big) \, 
q_C \, U_C^\alpha(\tau_C) \, d\tau_C  + . . . 
\label{eq:Green_fn_solution_4}
\end{multline}
and the divergence of the four-potential becomes
\begin{multline}
\frac{\partial A^\alpha (x)}{\partial x^\alpha} 
= - \int_{\tau_A^{(i)}}^{\tau_A^{(f)}} 
\Big[ \frac{d}{d\tau_A} \delta\Big( \big(x - x_A(\tau_A) \big)^2 \Big) \Big] \, q_A \, d\tau_A \\
- \int_{\tau_B^{(i)}}^{\tau_B^{(f)}} 
\Big[ \frac{d}{d\tau_B} \delta\Big( \big(x - x_B(\tau_B) \big)^2 \Big) \Big] \, q_B \, d\tau_B \\
- \int_{\tau_C^{(i)}}^{\tau_C^{(f)}} 
\Big[ \frac{d}{d\tau_C} \delta\Big( \big(x - x_C(\tau_C) \big)^2 \Big) \Big] \, q_C \, d\tau_C - . . . 
\label{eq:divergence_2}
\end{multline}

The integrals in equation (\ref{eq:divergence_2}) are grouped according to the Wheeler electron they belong to. 
For each Wheeler electron, the negative electric charges of its composing particles are replaced by $- e$, while the
positive electric charges are replaced by $e$, where $e$ is the electric charge of a proton. Then
the limits of integration for the positively charged particles are swapped, this introduces a change of sign.
The integrals for the negatively charged particles (the \lq\lq electrons\rq\rq) are calculated by going forward in time,
while the integrals for the positively charged particles (the \lq\lq positrons\rq\rq) are calculated by going backward in time.
(This is in complete agreement with Wheeler's observation that 
\lq\lq positrons could simply be represented as electrons going from the future to the past\rq\rq \cite{Feynman1965}.)
In this way, the contribution to equation (\ref{eq:divergence_2}) of each Wheeler electron is written as the closed path integral of a 
perfect differential, which is zero. For example, 
assuming that particles $A$ and $D$ have negative electric charge, 
the Wheeler electron in Figure \ref{fig:Wheeler_electron} gives 
\begin{multline}
e \int_{I_1}^{F_1} \Big[ \frac{d}{d\tau_A} \delta\Big( \big(x - x_A(\tau_A) \big)^2 \Big) \Big] \, d\tau_A 
+ e \int_{F_2}^{I_1} \Big[ \frac{d}{d\tau_B} \delta\Big( \big(x - x_B(\tau_B) \big)^2 \Big) \Big] \, d\tau_B \\
+ e \int_{F_1}^{I_2} \Big[ \frac{d}{d\tau_C} \delta\Big( \big(x - x_C(\tau_C) \big)^2 \Big) \Big] \, d\tau_C 
+ e \int_{I_2}^{F_2} \Big[ \frac{d}{d\tau_D} \delta\Big( \big(x - x_D(\tau_D) \big)^2 \Big) \Big] \, d\tau_D \\
= e \oint_W \Big[ \frac{d}{d\tau_W} \delta\Big( \big(x - x_W(\tau_W) \big)^2 \Big) \Big] \, d\tau_W = 0,
\label{eq:divergence_3}
\end{multline}
where $W$ stands for $A$, $B$, $C$, or $D$, depending on which section of the integral we are.
Alternatively, one could also reparameterize the whole closed path.

In this way the Lorenz gauge condition is recovered in a very natural and intuitive way,
without the need to extend the time coordinate (a real number) to the complex plane.
These two apparently independent concepts, the Lorenz gauge and the time symmetric
action-at-a-distance electrodynamics, are seen to reinforce each other in very harmonious ways.
From this point of view, out of all the possible linear combinations of retarded and advanced potentials,
the time symmetric electrodynamics (where the two potentials enter with equal weights) holds special status.
And similarly, out of all the possible electromagnetic gauges (Lorenz, Coulomb, etc. \cite{Jackson2001,Jackson2002}),
the Lorenz gauge holds special status. No wonder Jos\'{e}~A.~Heras and Guillermo Fern\'{a}ndez-Anaya \cite{Heras2010}
have concluded that the electromagnetic potentials in the Lorenz gauge could be considered physical quantities.

Our verification of the Lorenz gauge condition, based on the electric charge conservation 
during the creation and annihilation of point particles,
complements quite nicely an alternative calculation 
based on the electric charge conservation written in differential form. 
It is well known that, once the retarded Li\'{e}nard-Wiechert potentials are assumed, 
the Lorenz gauge condition is recovered as
a direct consequence of the equation of continuity \cite{Griffiths2012,Redzic2016}.
This also happens with the time symmetric four-potential (\ref{eq:four_potential}) 
\begin{multline}
\frac{\partial A^\alpha (x)}{\partial x^\alpha} 
= \frac{1}{c} \int \Big[ \frac{\partial}{\partial x^\alpha} \delta\big( (x - x')^2 \big) \Big] \, J^\alpha (x') \, d^4x' \\
= - \frac{1}{c} \int \Big[ \frac{\partial}{\partial x'^\alpha} \delta\big( (x - x')^2 \big) \Big] \, J^\alpha (x') \, d^4x' \\
= \frac{1}{c} \int \delta\big( (x - x')^2 \big) \, \frac{\partial J^\alpha (x')}{\partial x'^\alpha} \, d^4x' = 0,
\label{eq:divergence_4}
\end{multline}
where we have integrated by parts, assuming that the electric sources $J^\alpha$ are localized in space and time.

While this last calculation apparently avoids the decomposition into Wheeler electrons, 
this decomposition is in fact hidden inside the equation of continuity. 
By taking the divergence of the four-current density (\ref{eq:four_current_density_2}),
and by performing the same procedural steps as done for the divergence of the four-potential,
instead of the Lorenz gauge condition we now recover 
the equation of continuity.

The divergence of the four-current density (\ref{eq:four_current_density}) is
\begin{multline}
\frac{\partial J^\alpha (x)}{\partial x^\alpha} 
= \int \Big[ \frac{\partial}{\partial x^\alpha} 
\delta^{4D}\big( x - x_A(\tau_A) \big) \Big] \, q_A \, U_A^\alpha(\tau_A) \, c \, d\tau_A \\
= - \int \Big[ \frac{\partial}{\partial x_A^\alpha} 
\delta^{4D}\big( x - x_A(\tau_A) \big) \Big] \, q_A \, \frac{dx_A^\alpha}{d\tau_A} \, c \, d\tau_A \\
= - \int \Big[ \frac{d}{d\tau_A} 
\delta^{4D}\big( x - x_A(\tau_A) \big) \Big] \, q_A \, c \, d\tau_A,
\label{eq:divergence_J}
\end{multline}
and the divergence of the four-current density (\ref{eq:four_current_density_2}) is
\begin{multline}
\frac{\partial J^\alpha (x)}{\partial x^\alpha} 
= - \int_{\tau_A^{(i)}}^{\tau_A^{(f)}} 
\Big[ \frac{d}{d\tau_A} \delta^{4D}\big( x - x_A(\tau_A) \big) \Big] \, q_A \, c \, d\tau_A \\
- \int_{\tau_B^{(i)}}^{\tau_B^{(f)}} 
\Big[ \frac{d}{d\tau_B} \delta^{4D}\big( x - x_B(\tau_B) \big) \Big] \, q_B \, c \, d\tau_B \\
- \int_{\tau_C^{(i)}}^{\tau_C^{(f)}} 
\Big[ \frac{d}{d\tau_C} \delta^{4D}\big( x - x_C(\tau_C) \big) \Big] \, q_C \, c \, d\tau_C - . . . 
\label{eq:divergence_J_2}
\end{multline}
which is zero, since the contribution of each Wheeler electron is zero.

\section{Conclusions}

The Lorenz gauge condition and the equation of continuity are usually considered from a
local perspective. In this case, they are verified by direct differentiation
of the expressions of the (retarded)
Li\'{e}nard-Wiechert potentials, 
the electric charge density, and the electric current density, 
known in an infinitesimal neighborhood of a point.
However, the Lorenz gauge condition and the conservation of electric charge
can also be considered from a global perspective. In this case, they are verified
by integrating a perfect differential function along a closed path. 
While the expressions of these two perfect differential functions are different
in the two corresponding derivations, the closed path is the same. 
This common  path,
most clearly noticeable in the formulas (\ref{eq:four_current_density_2}) and (\ref{eq:Green_fn_solution_4})
that give the four-current density and the four-potential,
is the newly discovered connection between the Lorenz gauge condition
and the conservation of electric charge. 
We also remark that, since this closed path
(in fact a sum of very special closed paths, consisting of all the Wheeler electrons
in the universe) goes into the past as well as into the future, by necessity
we have to use time symmetric electrodynamics 
when verifying the Lorenz gauge condition in this manner.

\end{document}